\documentclass{article}  
\usepackage{breckenridge}
\usepackage{graphicx}
\frompage{000} \topage{000}                                              %|
\def\rightmark{Elliptic flow measurements from STAR}
\def\leftmark{Raimond Snellings}
 
\title{Elliptic flow measurements from STAR} 
\authors{
{Raimond Snellings$^{1}$ for the STAR Collaboration %
}\\[2.812mm]
{\normalsize
\hspace*{-8pt}$^1$ NIKHEF, \\ 
Kruislaan 409, 1098 SJ Amsterdam, The Netherlands\\
E-mail: Raimond.Snellings@nikhef.nl\\[0.2ex] 
}}
 
\abstract{In these proceedings some of the highlights of the elliptic
  flow measurements from STAR at
  $\sqrt{s_{_{NN}}} =$ 130 and 200 GeV for Au+Au collisions are presented.} 


\begin{document}
 
\maketitle
\setcounter{page}{1}

\section{Introduction}
\label{intro}

Elliptic flow characterizes the anisotropy in particle emission ``in''
and ``out'' of the reaction plane. The word flow is used to describe
collective behavior but does not necessarily imply a hydrodynamic
interpretation. 
Elliptic flow is commonly
characterized by the second harmonic coefficient $v_2$ of an azimuthal
Fourier decomposition of the the particle momentum distribution versus
the reaction plane. 

Based on general arguments it is thought that elliptic flow
develops mostly in the first few fm/$c$ ($<$ radius of the nucleus) and
thus provides information about the amount of thermalization achieved
{\it early} in the collision. In fact, the observed elliptic
flow for charged and identified particles at RHIC is interpreted as:
\begin{itemize}
\item 
one needs very strong interactions between the quarks and gluons at
very early times in the collision~\cite{McLerran}. 
\item
a well developed quark-gluon plasma~\cite{plasma}
\end{itemize}

\section{Integrated elliptic flow and non-flow contributions}
\label{nonflow}  
Experimentally the reaction plane is
not known, and elliptic flow is often reconstructed
from two-particle azimuthal correlations.
Two-particle azimuthal
correlations can be affected by many other sources besides elliptic
flow. These so called non-flow effects could be large~\cite{kovchegov}
and this would change the interpretation of
strong re-interactions of the constituents early in the collision. 
STAR has estimated the
contribution of non-flow in the first elliptic flow paper from
RHIC~\cite{starflow1}. 
\begin{figure}[htb]
  \begin{minipage}[t]{0.5\textwidth}
    \begin{center}
      \includegraphics[width=0.95\textwidth]{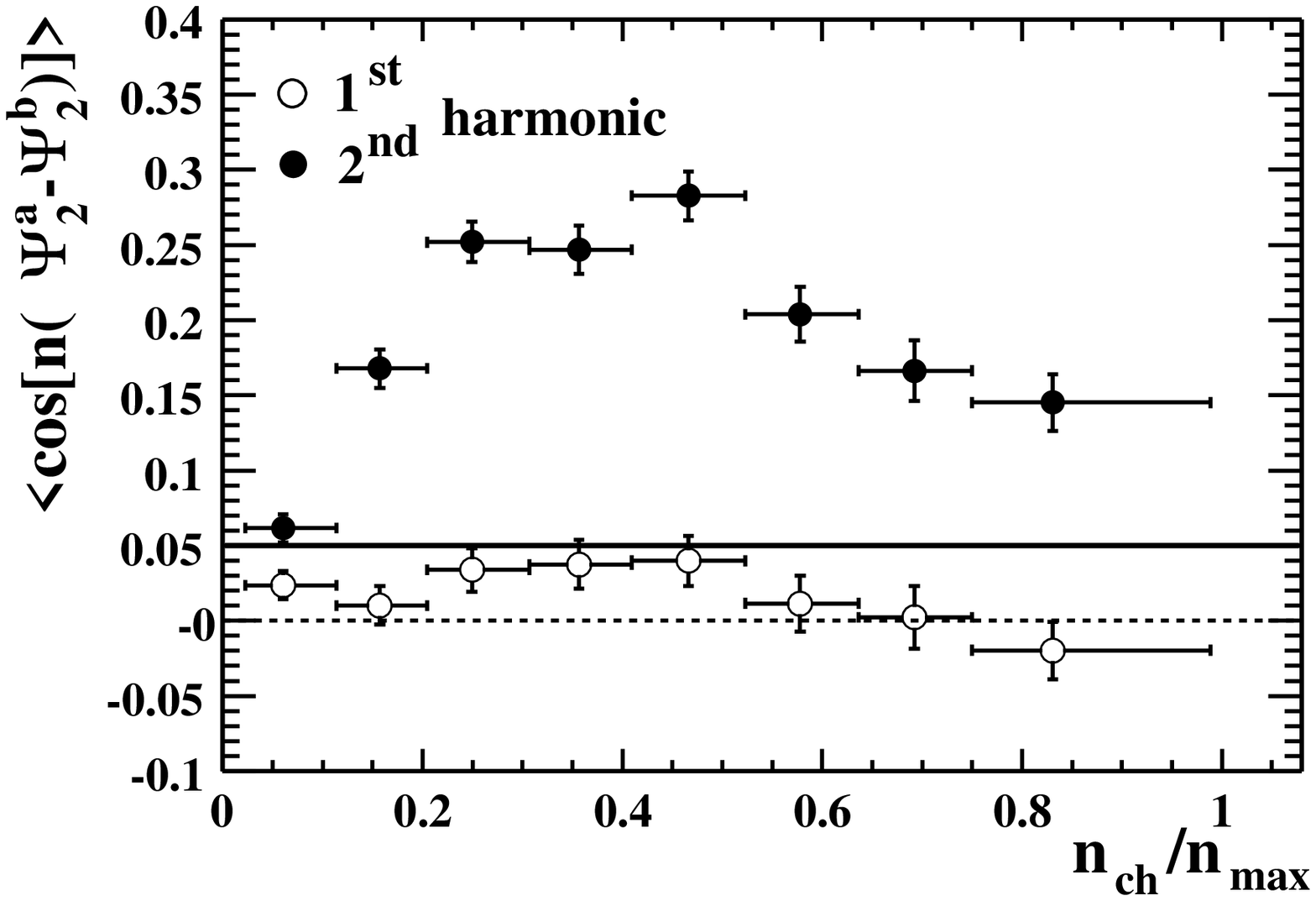}
      \caption{Correlation between the event plane angles determined
        from two independent subevents for the first and second harmonic.}
      \label{subevent}
    \end{center}
  \end{minipage}
  \hspace{\fill}
  \begin{minipage}[t]{0.5\textwidth}
    \begin{center}
      \includegraphics[width=0.95\textwidth]{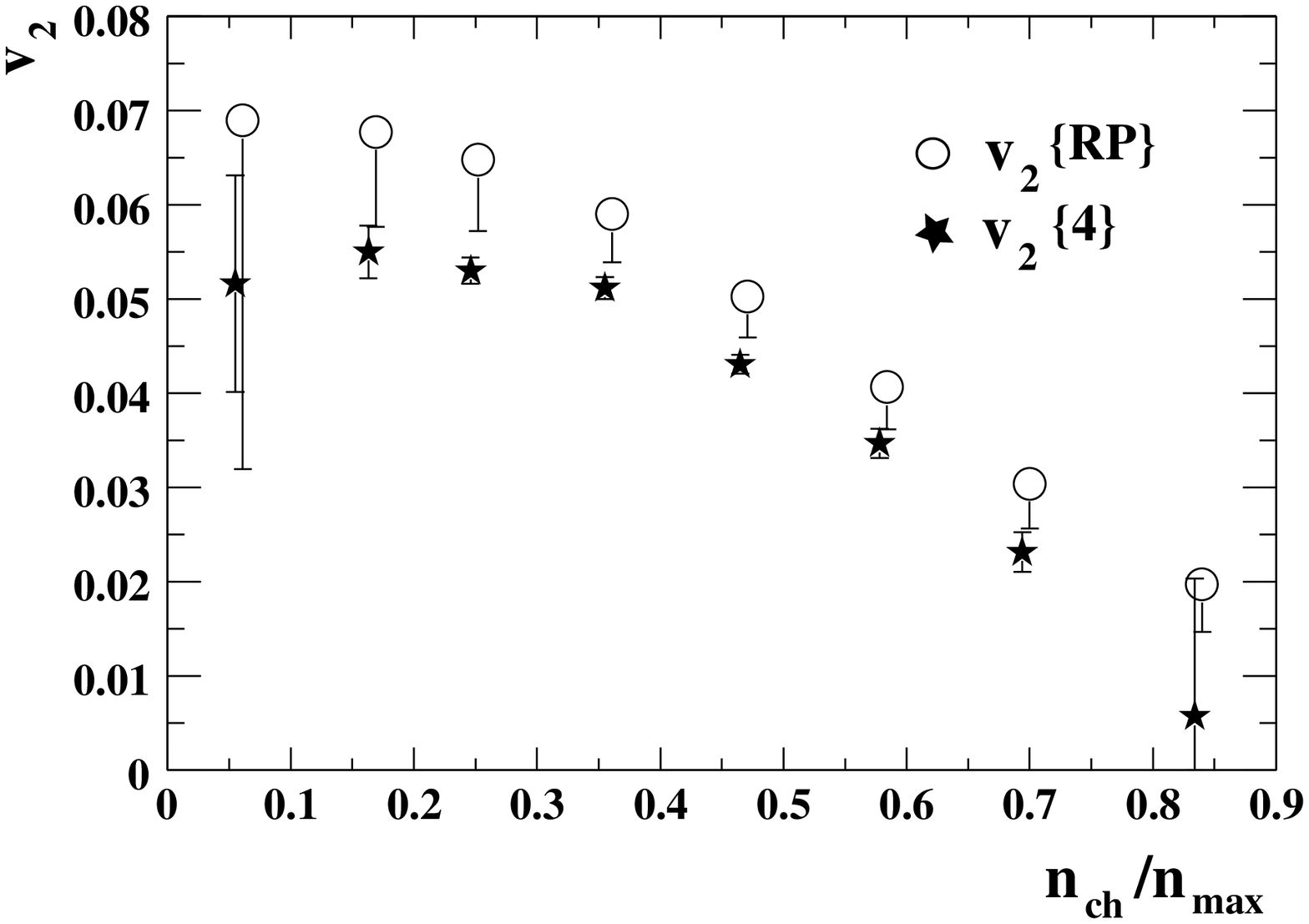}
      \caption{Identified particle elliptic flow versus centrality
        from the reaction plane, $v_2$\{RP\}, and the four
        particle cumulant, $v_2$\{4\}, method.}
      \label{cumulant}
    \end{center}
  \end{minipage}
\end{figure}
Figure~\ref{subevent} taken from~\cite{starflow1} shows the correlation
between two event plane angles for two independent subevents for the 
first and second harmonic as a function of centrality. 
In the case of flow the correlation of the
second harmonic event plane angles will be proportional to M$\cdot
v_2^2$, where M is the multiplicity of particles used in the determination
of the event plane. The observed peaked shaped as a function of
centrality is characteristic of elliptic flow. This peaked
shape originates from the fact that M increases as a function of
centrality while elliptic flow decreases. Every model description of a
heavy-ion collision which includes final state interactions will yield
such a shape, the magnitude and peak position as a function of
centrality will depend on the amount of re-interactions. Non-flow
contributions will be monotonic or almost constant for this quantity,
which is also true in the specific case of non-flow due to mini-jets as
calculated in~\cite{kovchegov}. Based on this correlation the
estimated maximum non-flow contribution, taken constant as a
function of centrality, was 0.05. 
The propagation of this estimate to the measured $v_2$ values is shown in
Fig.~\ref{cumulant}. In this figure the $v_2$ values versus
centrality are shown as open circles and the asymmetric uncertainties
are the non-flow estimates~\cite{starflowQM1}.

More recently a new analysis method based on cumulants was
proposed~\cite{olli_cumulants}
which utilizes the fact that true flow is a multi-particle correlation. 
The obtained elliptic flow values using this method for four particle
correlations, $v_2$\{4\}, is also shown in Fig.~\ref{cumulant} as solid stars. A
detailed description of the cumulant analysis in STAR for the
$\sqrt{s_{_{NN}}} = 130$ GeV data is given in Ref.~\cite{starflowPRC}.  
The reduction in the integrated elliptic flow between the two methods
shows the (possible) 
contribution of non-flow effects at this energy. There are however two
important caveats associated with calculating $v_2$ using two or
multi-particle correlations.
First the four particle cumulant is reliable when the
magnitude of $v_2$\{4\} $> 1/N^{3/4}$~\cite{olli_cumulants,kovchegov},
where N is the number of ``clusters'' contributing to the measurement. 
However, this is an important improvement over the two particle analysis
which is reliable when  $v_2$\{2\} $> 1/N^{1/2}$.
Secondly, event by event fluctuations could affect $v_2$\{2\}
differently than $v_2$\{4\}.
The two particle analysis is sensitive to $v_2^2$
while the four particle cumulant analysis is sensitive to the difference between
$v_2^4$ and $v_2^2$. The $v_2$ value is obtained by averaging over
events and due to event by event fluctuations in general 
$\langle v_2 \rangle^2 \ne \langle v_2^2 \rangle$ 
and $\langle v_2 \rangle^4 \ne \langle v_2^4 \rangle$. This 
can also lead to a reduction in the $v_2$ obtained from the four
particle compared to the two particle correlation methods~\cite{starflowPRC,voloshinQM2}.

\begin{figure}[htb]
  \begin{minipage}[t]{0.51\textwidth}
    \begin{center}
      \includegraphics[width=0.95\textwidth]{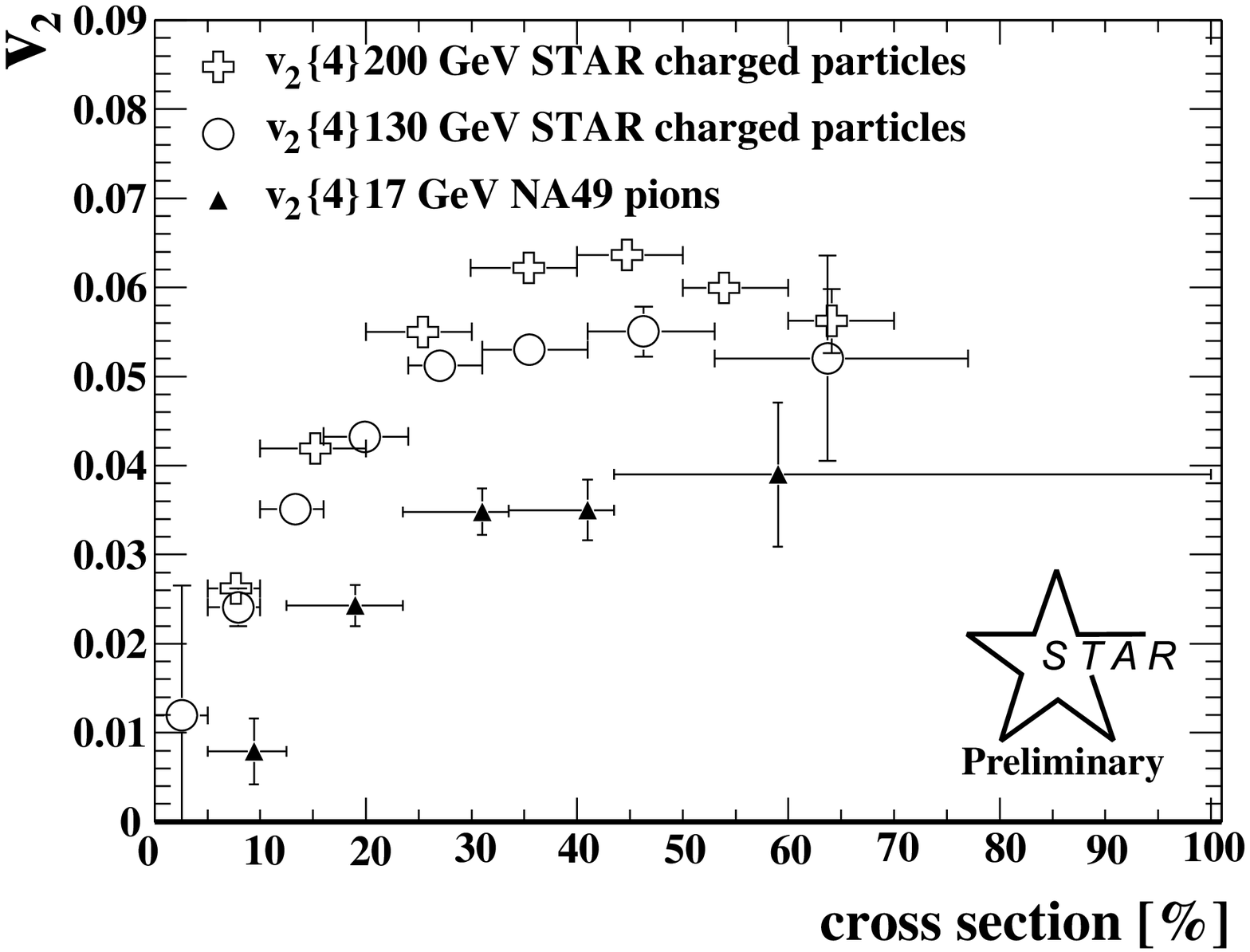}
      \caption{Centrality dependence of elliptic flow 
        for collisions at $\sqrt{s_{_{NN}}} =$ 17, 130 and 200 GeV.}
      \label{energy_centrality}
    \end{center}
  \end{minipage}
  \hspace{\fill}
  \begin{minipage}[t]{0.49\textwidth}
    \begin{center}
      \includegraphics[width=0.95\textwidth]{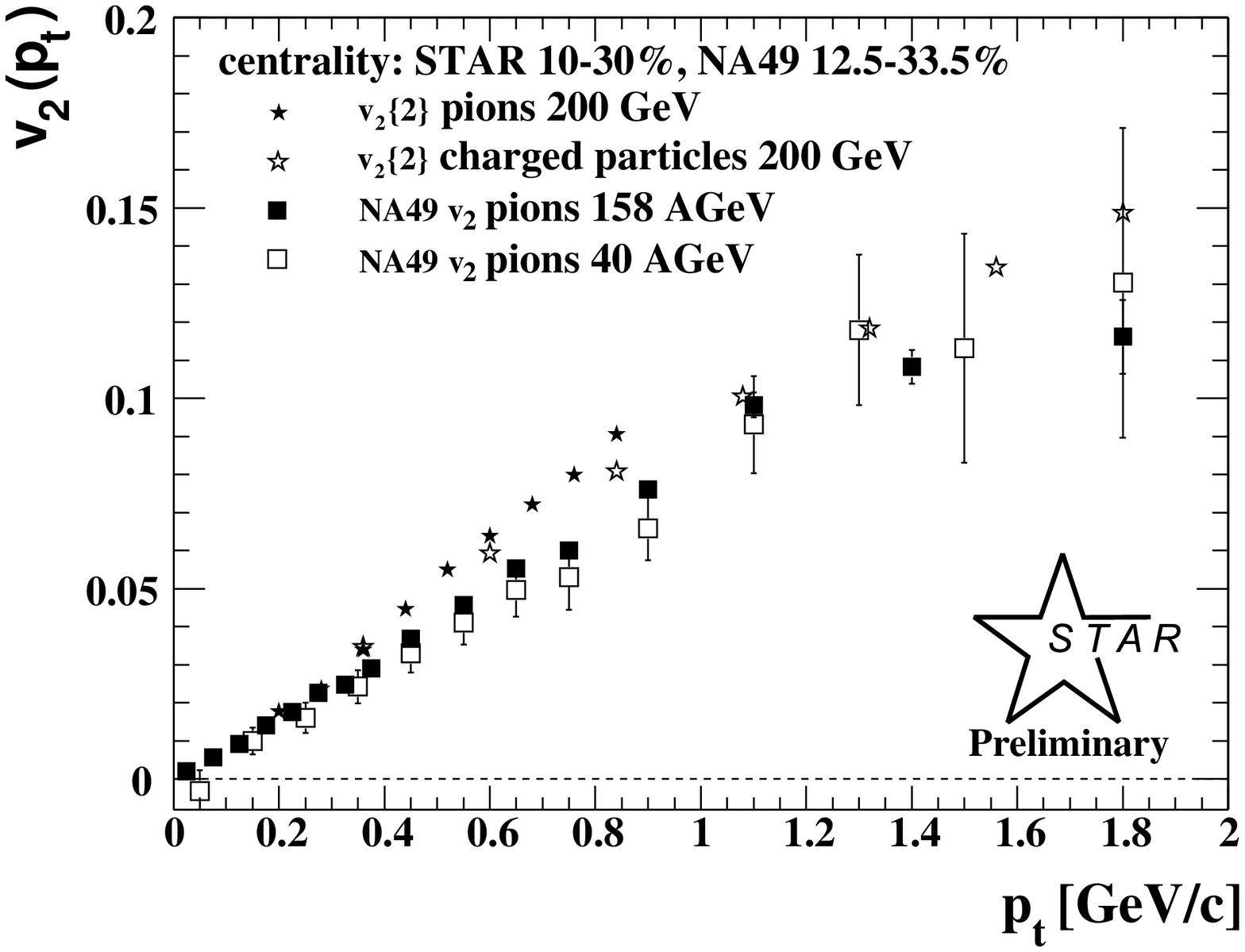}
      \caption{Comparison of $v_2$($p_t$) for SPS and RHIC energies.}
      \label{STAR_NA49}
    \end{center}
  \end{minipage}
\end{figure}
Even with these caveats, the most reliable way of
calculating $v_2$ when the reaction plane is not known is via a
higher order cumulant method. Figure~\ref{energy_centrality} shows the measured
$v_2$\{4\} values as a function of centrality for $\sqrt{s_{_{NN}}} =$
130 and 200 GeV. The elliptic flow as a function of centrality for
both energies is very similar. The maximum difference is about 20\%,
however note that only the statistical uncertainties are shown. For the
preliminary 200 GeV results the systematic uncertainties are also
about 20\%. Also shown are the measurements from NA49 at the CERN
SPS. The increase in the
integrated elliptic flow between SPS and RHIC energies can be caused by
an increase in the mean transverse momentum or due to a higher slope
of the differential $v_2$($p_t$)~\cite{voloshinQM2}. 
Figure~\ref{STAR_NA49} shows the
differential $v_2$($p_t$) for SPS and RHIC energies. 
While the slope increases, which indicates the increase in
elliptic flow at RHIC, the dominant contribution to the integrated
elliptic flow comes from the value of $v_2$ around mean $p_t$ of the particles. 
In the $p_t$ range of $350 - 500$ MeV/$c$, the $v_2$($p_t$) values at SPS and
RHIC energies are very close. However the mean $p_t$ difference between
pions at the SPS and charged particles at RHIC is about 150 MeV/$c$
which already accounts for most of the difference in integrated $v_2$.

\section{Elliptic flow for identified particles}
\label{flowpid}

\begin{figure}[htb]
  \begin{minipage}[t]{0.5\textwidth}
    \begin{center}
      \includegraphics[width=0.95\textwidth]{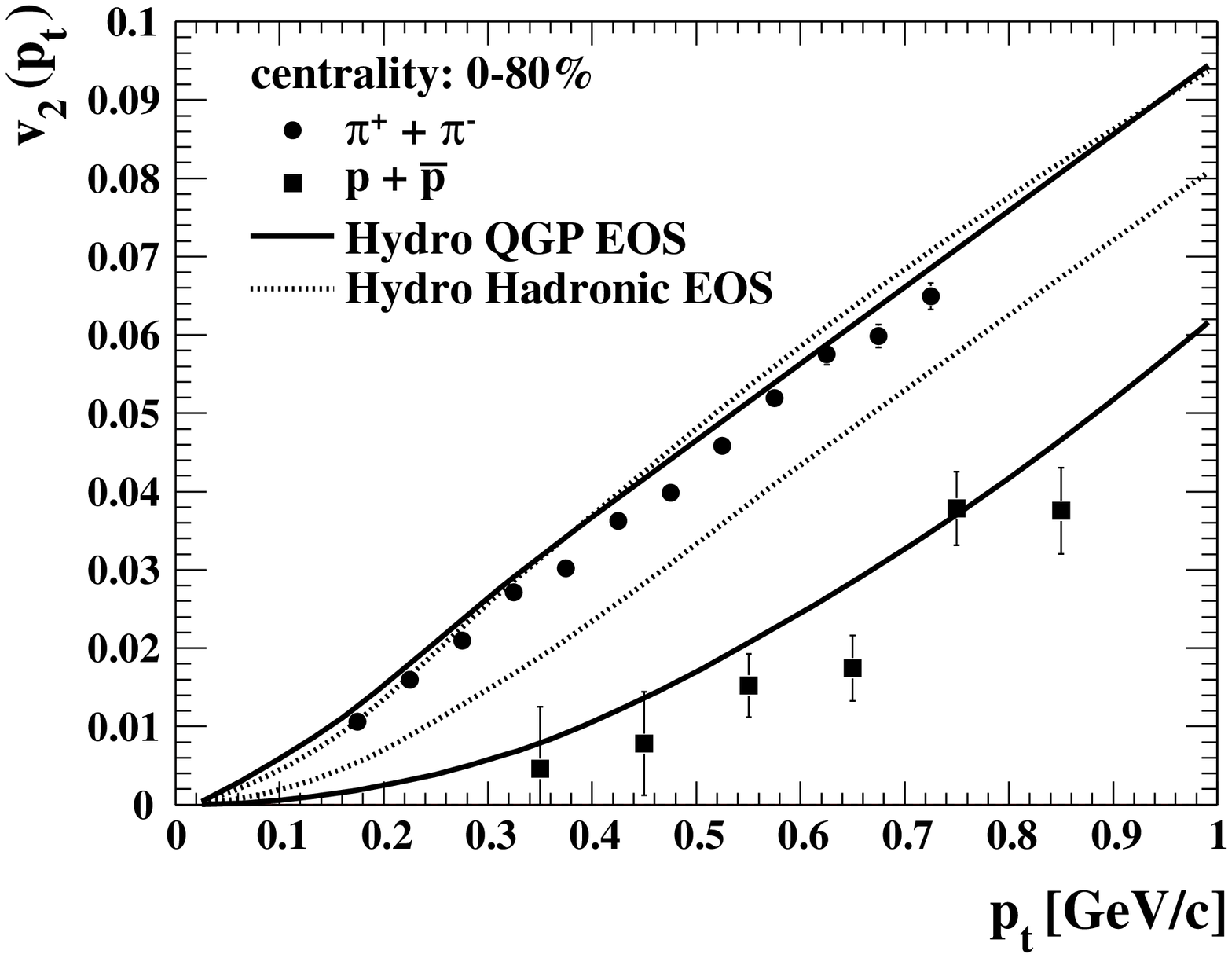}
      \caption{$v_2$($p_t$) for pions and protons at $\sqrt{s_{_{NN}}}
        =$ 130. The lines are hydrodynamical model calculations.}
      \label{Hydro_EOS}
    \end{center}
  \end{minipage}
  \hspace{\fill}
  \begin{minipage}[t]{0.5\textwidth}
    \begin{center}
      \includegraphics[width=0.95\textwidth]{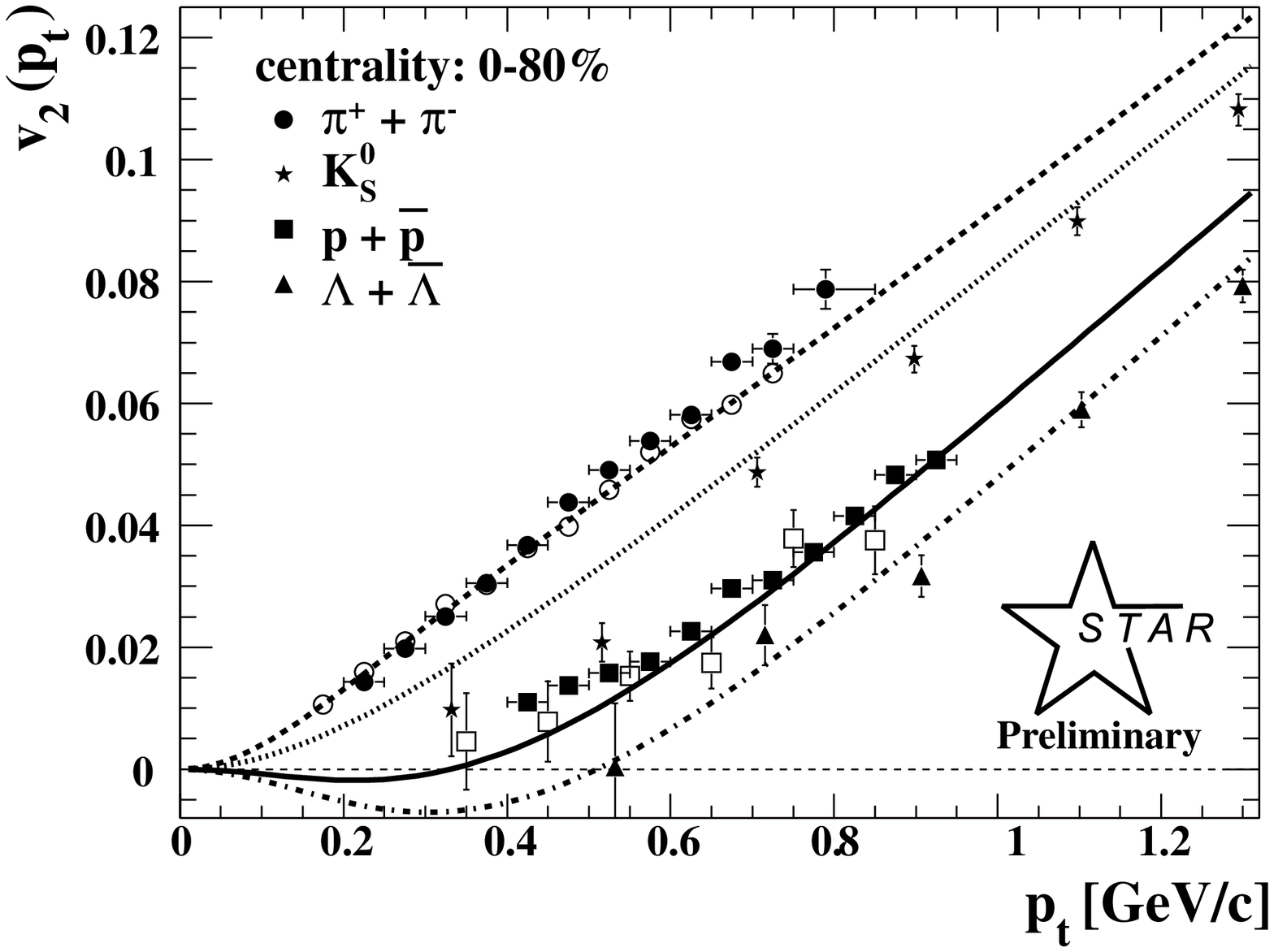}
      \caption{Comparison of $v_2$($p_t$) for identified particles at 
        $\sqrt{s_{_{NN}}} =$ 130 and
        200 GeV Au+Au collisions.} 
      \label{pid130_200}
    \end{center}
  \end{minipage}
\end{figure}
The elliptic flow as a function of transverse momentum, $v_2$($p_t$), 
depends on the temperature, radial flow velocity, azimuthal variation of the
transverse flow velocity and the spatial anisotropy of the system at
freeze-out~\cite{starflowpid}. 
The measurement of $v_2$($p_t$) for different particle
masses constrains these parameters with reasonable precision~\cite{starflowpid}. 
These freeze-out parameters in combination with the initial conditions
strongly constrain the Equation Of State (EOS).
In Fig.~\ref{Hydro_EOS} the measured $v_2\{2\}$ versus $p_t$ at $\sqrt{s_{_{NN}}}
=$ 130 GeV for pions and protons + antiprotons is
shown~\cite{starflowpid} together with hydrodynamical model
predictions~\cite{pasi} for two different EOS. 
This clearly illustrates that the
heavier particles are more sensitive to the underlying EOS. This can
be understood from the fact that the lighter particles are very
sensitive to the freeze-out temperature while the heavier particles
more directly reflect the flow. From Fig.~\ref{Hydro_EOS} it is clear
that for these model calculations the data prefer the quark-gluon
plasma EOS. 
Figure~\ref{pid130_200} shows in solid symbols the $v_2\{2\}$ versus $p_t$ for pions,
$K_S^0$'s, protons + antiprotons and $\Lambda + \bar{\Lambda}$ at
$\sqrt{s_{_{NN}}} =$ 200 GeV. For comparison the $v_2$($p_t$) for
pions and protons + antiprotons at $\sqrt{s_{_{NN}}} =$ 130 GeV in
open symbols are included. 
The lines shown are the results from the blastwave fits to the $\sqrt{s_{_{NN}}}
=$ 130 GeV data~\cite{starflowpid}.
From this comparison it is clear that the
$v_2$($p_t$) for different particles at these different energies is
very similar. The pions do show however a slightly higher slope at $\sqrt{s_{_{NN}}}
=$ 200 GeV.

\section{Elliptic flow at intermediate transverse momentum}
\label{flowhighpt}
Elliptic flow measurements can quantify the
possible modifications of the created medium on the particle
yields as a function of $p_t$. A medium modification like the
predicted mechanism of parton
energy loss, jet quenching,
will inevitably lead to a finite $v_2$ at high-$p_t$.

\begin{figure}[htb]
  \begin{minipage}[t]{0.51\textwidth}
    \begin{center}
      \includegraphics[width=0.95\textwidth]{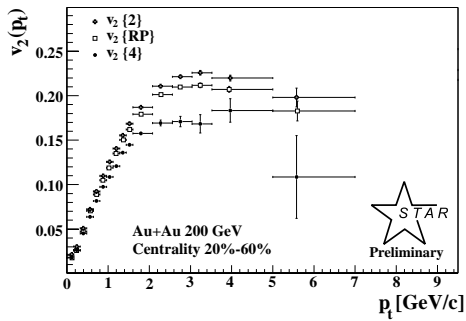}
      \caption{Charged particle elliptic flow obtained by two
        ($v_2$\{2\} and $v_2$\{RP\}) and four ($v_2$\{4\}) particle
        correlation methods.}
      \label{charged_v2_highpt}
    \end{center}
  \end{minipage}
  \hspace{\fill}
  \begin{minipage}[t]{0.49\textwidth}
    \begin{center}
      \includegraphics[width=0.95\textwidth]{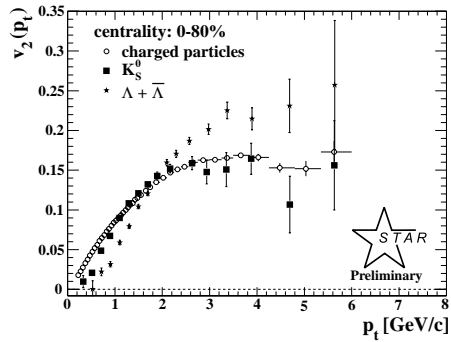}
      \caption{Elliptic flow versus $p_t$ for charged particles,
        $\Lambda + \bar{\Lambda}$ and $K_S^0$.} 
      \label{pid_v2_highpt}
    \end{center}
  \end{minipage}
\end{figure}
Figure~\ref{charged_v2_highpt} shows the measured $v_2$($p_t$) up to 7
GeV/$c$ obtained by two, $v_2$\{2\} and $v_2$\{RP\}, and four,
$v_2$\{4\}, particle correlation methods. The measurement of $v_2$\{4\} 
shows that there is a significant amount
of elliptic flow out to $p_t = 5-6$ GeV/$c$.
The medium modification inferred
from this observable in addition to
the suppression of the single inclusive particle yield~\cite{dunlop},
characterized by R$_{AA}$, and the
disappearance of high-$p_t$ angular back to back correlations,
as seen in the centrality dependence of I$_{AA}$~\cite{harris}, is in
qualitative agreement with the predictions of jet quenching.   
At high-$p_t$ the measurement of R$_{AA}$, I$_{AA}$
and $v_2$ for the different particle species can provide a better
constraint on the underlying mechanism responsible for the
modification of the particle yield. 
That R$_{AA}$ depends on particle species was observed by
PHENIX~\cite{phenix_RAA_pid} for $\pi^0$'s and charged particles.
The comparison of the single
inclusive particle yield for $\Lambda + \bar{\Lambda}$ and $K_S^0$ in
central and peripheral collisions as measured by STAR also shows this
species dependence~\cite{belwied}. 
The suppression in the
single inclusive particle yield at intermediate $p_t$ for $K_S^0$ is
stronger than for $\Lambda + \bar{\Lambda}$. 
The measured $v_2\{2\}$, shown in Fig.~\ref{pid_v2_highpt}, 
for $\Lambda + \bar{\Lambda}$ and $K_S^0$
at the same intermediate $p_t$ is also particle dependent. The
$v_2$($p_t$) above 2 GeV/$c$ for the $\Lambda + \bar{\Lambda}$ is
larger than the $v_2$($p_t$) of $K_S^0$. 
It is assumed that the non-flow contributions in this measurement are
approximately equal for both particle species and
therefore the difference is a real difference in flow.

Explaining the origin of this suppression of single inclusive particle
yield and elliptic flow for identified particles at intermediate-$p_t$
due to jet quenching alone seems not to work. To understand this
observed behavior better a measurement of the contribution of the
initial state Cronin effect
in dA as well as a measurement of I$_{AA}$ for 
$\Lambda + \bar{\Lambda}$ and for $K^0_S$ is needed. 
Another possible explanation is that the hydro-like behavior 
extends further in $p_t$ for $\Lambda + \bar{\Lambda}$ then for $K^0_S$.  

Another puzzle is the rather large values of elliptic flow at
intermediate $p_t$. The interpretation that this is caused by radiative
energy loss or due to inelastic interactions in a parton cascade would
lead to rather large initial parton densities. 
Corrections for non-flow would already
reduce the elliptic flow values and therefore the parton densities
needed. In addition a mechanism like 
parton coalescence would require much smaller parton elliptic flow to
account for the measured particle elliptic flow and would also explain the
different $v_2$ values observed at intermediate-$p_t$ for $\Lambda +
\bar{\Lambda}$ and $K^0_S$~\cite{molnar}.

\section{Conclusions}
\label{concl}

Large elliptic flow values have been measured by STAR both at
$\sqrt{s_{_{NN}}} =$ 130 and 200 GeV Au+Au collisions. 
The large magnitude of the charged and identified particle elliptic flow at
low-$p_t$ is interpreted as due to strong interactions between the
partons at early times in collision and even approaches the ideal
hydrodynamical limit.
While non-flow is not negligible at higher-$p_t$, elliptic flow
extends at least up to 6 GeV/$c$. 
This unambiguously shows the effect of medium
modification on the particle yield in this transverse momentum
range. The dependence of the elliptic flow and the 
single particle inclusive yield on particle species at
intermediate-$p_t$ shows that radiative energy loss can not be the
only medium induced modification of the particle yield in this $p_t$-range.

%\vfill\eject
\end{document}